\documentclass[%
 aip,
 amsmath,amssymb,
 reprint,%
]{revtex4-1}
\usepackage{graphicx}
\usepackage{amsmath,amssymb}
\usepackage{color}
\usepackage{textcomp}
\usepackage{float}

\begin{document}

\begin{titlepage}
\title{Stress-stress Correlations Reveal Force Chains in Gels}
\author{H. A. Vinutha}
\author{Fabiola Diaz Ruiz}
\affiliation{Department of Physics, Institute for Soft Matter Synthesis and Metrology, Georgetown University, Washington, DC, USA}
\author{Xiaoming Mao}
\affiliation{Department of Physics, University of Michigan, Ann Arbor MI, USA}
\author{Bulbul Chakraborty}
\affiliation{Martin Fisher School of Physics, Brandeis University, Waltham MA, USA}
\author{Emanuela Del Gado}
\affiliation{Department of Physics, Institute for Soft Matter Synthesis and Metrology, Georgetown University, Washington, DC, USA}

% word limit 600; currently 292
\begin{abstract}
{
We investigate the spatial correlations of microscopic stresses in soft particulate gels, using 2D and 3D numerical simulations. We use a recently developed theoretical framework predicting the analytical form of stress-stress correlations in amorphous assemblies of athermal grains that acquire rigidity under an external load. These correlations exhibit a pinch-point singularity in Fourier space leading to long-range correlations and strong anisotropy in real space, which are at the origin of force-chains in granular solids. Our analysis for the model particulate gels at low particle volume fractions demonstrates that stress-stress correlations in these soft materials have characteristics very similar to those in granular solids and can be used to identify force chains. We show that the stress-stress correlations can distinguish floppy from rigid gel networks and the intensity patterns reflect changes in shear moduli and network topology, due to the emergence of rigid structures during solidification.}
\end{abstract}
\maketitle
\end{titlepage}
%Intro 
\section{Introduction}
Soft particulate gels consist of particulate matter (polymers, colloids, proteins etc) aggregated in a solid matrix, which is embedded in a fluid and typically sparse and porous \cite{DelGadoBookChapter2016}. Gels of this type are 
found in the body tissues, in food, drugs, personal care products, and even in cement used in concrete based construction \cite{petekidis_wagner_2021,Ioannidou2016}. Their structures can be very soft and re-configurable and their stress response is determined by a complex interplay between molecular cohesion or surface interactions, microstructural reorganization and external driving. As a consequence, understanding, predicting and designing the mechanics and rheology of these materials remains very challenging. 

While particulate gels can be extremely soft because of the overall low solid content, the strength of the interactions that drive the aggregation of the colloidal units into the gel network must be large enough with respect to $k_{B}T$ to overcome thermal fluctuations and stabilize the gel structural elements and their connections \cite{anderson2002insights,royall2021real}. The implication is that the elasticity is largely enthalpic in nature in these gels, contrarily to polymer networks, and that the interaction strength, together with the total volume fraction of the solid content, controls elastic moduli, viscoelastic spectra, and nonlinear response. Remarkably, the interaction strength alone does not allow to predict the gel properties: the morphology of the gel networks, the structural elements and their connectivity, may change a lot due to different kinetics at play during the gel self-assembly, leading to huge variations of the mechanical response, sometimes even for identical compositions of the initial particle solution. 

The solidification processes through which particulate gels form, initiated, for example, via irreversible fractal aggregation or equilibrium phase separation, are typically sources of frozen-in stresses \cite{bouzid2017elastically} and help build a memory in their history dependent response \cite{zhang2021soft}: they determine how local stresses and mechanical heterogeneities get embedded and remain frozen-in, in the gel structure, during gelation. The rigidity attained ultimately depends on the geometry of the microstructure that self-assembles during solidification to accommodate the local and global constraints of mechanical equilibrium. Consequently, significant structural correlations, induced for example by attractive interactions, are naturally incorporated in the way stresses are transmitted through the rigid backbone of the final solid. This process of self-organization can provide an explanation for the possible emergence of rigidity at very low densities in colloidal gels, since the structural correlations are precisely those that satisfy the constraints of mechanical equilibrium, as indeed shown in Ref.\cite{zhang2019correlated}. The fact that this self-organization occurs in overall amorphous and spatially heterogeneous structures also naturally leads to stress localization which manifests itself clearly under deformation \cite{leocmach2014creep,colombo2014stress,filiberti2019multiscale,aime2018microscopic}. 

The tendency to localize strain and stresses and the signature of mechanical heterogeneities, shown under deformation, may suggest that stress transmission could be very anisotropic and strongly localized even in the material at rest \cite{alexander1998amorphous}, just as a result of the microstructural development during gelation. As a consequence, phenomena such as force chains in granular solid, where forces are transmitted across the sample along lines of grains forming a sparse network \cite{cates1998jamming,radjai1998bimodal,gendelman2016determines,vinutha2019force,wang2020connecting}, may be relevant to the physics of soft particulate gels as well. 
Detecting force chains and identifying the stress bearing part of gel structures is, however, incredibly difficult in experiments and elusive even in numerical simulations of model materials, since it requires disentangling mechanical heterogeneities from structural heterogeneities and relies upon threshold values for local stresses or forces which are not easily justified. For amorphous solids, recent theoretical studies have clarified that just the constraints of mechanical equilibrium and of structural isotropy guarantee that the spatial correlations of the stresses are long-ranged \cite{henkes2009statistical,mao2009soft,degiuli2018field,lemaitre2018stress}. The emergence of elasticity can then be obtained, even in granular solids where the grain contacts cannot be regarded as elastic springs, within a new theoretical framework that can be mapped onto a tensorial electromagnetism with vector charges (VCT) \cite{pretko2017generalized,nampoothiri2020emergent,nampoothiri2022tensor}. In particular, this VCT framework provides analytical predictions for the stress-stress correlations, demonstrating that they are not only long-ranged but also anisotropic, and clearly identifying the presence of force chains in disordered assemblies of grains which are jammed under an external load.   

Here we use the predictions of the VCT framework to compute stress-stress correlations in model particulate gels, obtained in 2D and 3D numerical simulations. We find that the intensity patterns of these correlations in Fourier and real space share several common features with those detected in experiment and simulations of granular solids, and predicted by the VCT framework \cite{nampoothiri2020emergent,nampoothiri2022tensor}. In particular, we recover the anisotropy of the stress-stress correlations, which therefore becomes an indicator of the presence of force chains in particulate gels. Moreover, we show that the stress-stress correlations distinguish floppy from rigid gels and are sensitive to the distance of the model gels from the rigidity threshold. These results pave the way to develop further understanding of stress transmission in soft materials and provide the first basis to develop a VCT framework for soft gel materials. 

The paper is organized as follows. In section \ref{theory-simulations}.A we briefly review the basic ingredients of the VCT framework and its predictions for the stress-stress correlations in granular solids, while section \ref{theory-simulations}.B contains the information on the numerical models and simulations used, and in section \ref{theory-simulations}.C the calculations of the stress correlations are described. We then discuss the results obtained in section \ref{results} and provide a summary and outlook in section \ref{conclu}. 
% What we do 
\section{Theory predictions and numerical simulations}
\label{theory-simulations}
\subsection{Elasticity theory of athermal amorphous solids and VCT framework.}\label{theorypred}
At the heart of the mechanical response of jammed amorphous solids is the athermal nature of the force-bearing networks that obey the constraints of mechanical equilibrium, implemented locally: each particle satisfies the constraints of force and torque balance. In a continuum formulation the force-balance  constraints are expressed as :
\begin{equation}\label{forcebalance}
\partial_i \sigma_{ij} = f_{j}
\end{equation}
and torque balance leads to the symmetry of the stress tensor.
However, in $D$ dimensions, $d$ conditions of mechanical equilibrium are not enough to solve for the $D(D+1)/2$ components of the stress tensor. Canonical continuum elasticity solves for the stress components by defining the strain in terms of the displacement field from the crystal reference structure, which is uniquely defined, and uses the constitutive relation of linear elasticity to obtain a complete set of equations \cite{timoshenko1951theory,landau2012theory}. In the case of amorphous solids, there is no unique reference structure to define strain and the continuum elasticity description does not work. The VCT framework described in Ref. \cite{nampoothiri2020emergent} provides a stress-only framework to describe elasticity in such solids. This stress-only theory of elasticity is defined by equations that bear a remarkable similarity to that of classical elasticity theory\cite{nampoothiri2022tensor}:%{\color{red} BC:  I have taken the liberty of inserting the discussion below
\begin{align}
     \partial_i \sigma_{ij} &= f_j^{\rm~ external},\nonumber\\
     E_{ij}=\frac{1}{2}(\partial_i\varphi_j+\partial_j\varphi_i) &\implies \epsilon_{iab} \epsilon_{jcd} \partial_a \partial_c E_{bd}=0,\nonumber\\
     \sigma_{ij} = (\delta_{ijkl} + \chi_{ijkl})& E_{kl} \equiv {\Lambda^{-1}_{ijkl}E_{kl}.}
    \label{eq:VCTG_field_eqns}
\end{align}
% {\color{red}Eq. 29 in our long paper on the arxiv.}
Here, $\hat{\sigma}$ is the stress-tensor field, which is related to $\hat{E}$ via an emergent elastic modulus tensor, 
$\hat{\Lambda}^{-1}$.  $\hat{E}$ plays a role analogous to the strain tensor in classical elasticity theory, and external, body forces such as gravity are represented by $\vec{f}$. The two crucial differences from classical elasticity theory are that (a) the physical displacement field defining strain is replaced by the $\varphi$ field, which  is a gauge potential  since there is no unique reference structure, and (b) the $\Lambda^-1$ tensor is an emergent elastic modulus tensor that reflects the coarse-grained properties of the self-assembled, force-bearing network and specifically the force-dipoles that represent the frustration in the interactions between particles.%}

The VCT framework makes explicit predictions about stress-stress correlations and stress response in terms of the emergent elastic modulus tensor, $\Lambda^{-1}$. Here we summarize crucial features of the stress-stress correlations. %provides a continuum description of the disorder-averaged response of stresses in amorphous solids. Here, we summarize a few crucial findings of this study when applied to granular materials. 
[i] A hallmark of the theory is the appearance of a characteristic pinch-point in the stress-stress correlations, $C_{ijkl}(q) =  \langle \sigma_{ij}(q) \sigma_{kl}(-q) \rangle$. These predictions have been tested against experimental measurements in frictional granular materials, and in 2$D$ and in 3$D$ model assemblies of frictionless soft grains. [ii] The chain-like structures, commonly referred to as force-chains, are a visual representation of the the highly anisotropic nature of $C_{ijkl}$. [iii] The $q-$space correlations predicted by the theory, and observed in experiments imply a power-law decay at large lengthscales: $\langle \sigma_{ij}(r) \sigma_{ij}(0)\rangle \propto 1/r^D$, where the plus sign appears for longitudinal correlations and the minus sign for transverse correlations, and $D$ is the spatial dimension. 
In this paper we analyze stress-correlations in numerically simulated 2$D$ and 3$D$ particle gels, using the VCT framework.

\subsection{Simulation Details}
Gel configurations in two and three dimensions are obtained using models of $N$ interacting colloidal particles of diameter $d$ that undergo gelation as described in Refs.\cite{colombo2014self,colombo2014stress, bouzid2018network,bantawa2021microscopic,zhang2019correlated}. We perform Molecular Dynamics (MD) simulations in a cubic (square in 2D) box of size $L$ with periodic boundary conditions. We first reach thermal equilibrium at high temperature and then slowly quench the particle configurations to different target temperatures $T$, i.e., with different cooling rates $C_{r}$, using a Nos\'{e}--Hoover thermostat and allowing the particles to aggregate and form gel networks \cite{noro2000extended,colombo2014self}. The mechanical equilibrium in the final gel states is obtained by slowly drawing all kinetic energies with an overdamped dynamics, following the protocol described in Refs. \cite{zhang2019correlated,colombo2014stress,bantawa2021microscopic}. 
All the simulations are performed using LAMMPS \cite{lammps} modified to include specific particle interactions. We define an approximate volume fraction $\phi = \frac{\pi d^2 N}{4L^2}$ and $\phi = \frac{\pi d^3 N}{6L^3}$ in two and three dimensions, respectively.

For 2$D$ gels, we consider $N=10000$ monodisperse particles, interacting via a short range attractive potential 
$%$\begin{equation}
U(r) = A \epsilon \left(a\left( d/r \right)^{18} - \left(d/r\right)^{16}\right)$   
%\end{equation}
where $r$ is the interparticle distance, $A$ and $a$ are dimensionless constants. The values of $A=6.27$ and $a=0.85$ are chosen to obtain a short-ranged attractive well of depth $\epsilon$ and range $r_c \approx 0.3 d$ (the potential is cut and shifted to 0 at large distances). In 3D, we consider $N \approx 16000$ monodisperse particles that interact via the same short range attractive interactions and with an angular term that introduces bending rigidity to inter-particle bonds \cite{colombo2014stress,bouzid2018network,bantawa2021microscopic}.  

All the simulations quantities described in the following are expressed in the reduced units: $d$ for the unit of length, the unit of energy is $\epsilon$, the particle mass $m$ is the unit of mass. The reduced temperature is expressed in units $\epsilon/k_B$, where $k_B$ is the Boltzmann's constant. The unit of stress is therefore $\epsilon/d^3$.
When performing the quenches into aggregation from high temperature, the cooling rate $C_{\text{r}}$ is defined as  $\Delta T/\Delta t$, where $\Delta T$ is the distance between the initial and target temperatures in the gelation protocol described above and $\Delta t$ is the duration of the quench, with the unit time $\tau_{0}$ being $d\sqrt{m/\epsilon}$.   
In 2D, we show data for $\phi=0.5$ at $T=0.3$ (floppy gels) and $T=0.18$ (rigid gels). In 3D, we analyze gels obtained using the cooling rates $C_r = 9 \times 10^{-5} \epsilon/(\tau_{0} k_{B})$, and $9 \times 10^{-2}\epsilon/(\tau_{0} k_{B})$ at $\phi=0.125,0.075$ and $0.05$.

To infer rigidity, we use the pebble game algorithm on the 2D gel networks to identify rigid clusters and floppy regions \cite{jacobs1997algorithm,zhang2019correlated}: a rigid gel is characterized by a spanning rigid cluster. In 3D, we have instead computed the linear viscoelastic response of the gels and obtained their low frequency moduli \cite{colombo2014self,bouzid2018network,bouzid2018computing}. For rigid gels, the low frequency storage modulus is larger than the loss component. 

All the data for the stress correlations have been averaged over $10$ independently generated samples. 

\subsection{Computing stress correlation functions}
To compute the stress correlations in the gel samples, we coarse-grain the stress fluctuations in Fourier space by imposing a cutoff at large $q$, corresponding to $q_{max} = 2\pi/d$, i.e. we do not consider any stress fluctuations occurring at length scales shorter than $d$. The lower $q$ cutoff in Fourier space is set by the simulation box size $q_{min} = 2\pi/L$ and the periodic boundary conditions.
For every particle $i$ we compute the force-moment tensor defined as
\begin{equation}\label{FMpar}
\hat{\sigma}_i = \sum_{j=1}^{N_c} \vec{r}_{ij} \otimes \vec{f}_{ij}, 
\end{equation} 
where $\vec{r}_{ij}$ is the vector connecting the center of particles $i$ and $j$, and $\vec{f}_{ij}$ is the force between them. $N_c$ is the number of neighbors $j$ of particle $i$ within the interaction range  $r_{ij} \le r_c$.
%{\color{red} and $V_i$ is the volume (area) of grain $i$}. 
% We define a coarse-grained stress tensor from the Virial formula 
%{\color{red} BC Comment: is it possible to also look at the bond-based virial stress ? That is for each bond $ij$ ?} 
%The following equation does not seem to be needed.
%\[
 %   \hat{\sigma}(\vec{r}) = \frac{1}{V}  \sum_{i=1}^{N} \sum_{j \neq i, j=1}^{N_c} \vec{r}_{ij} \otimes \vec{f}_{ij},
%\]where $V$ is the coarse-grained volume or area.
%{\color{red} There are some factors of area and $\pi$ that are needed to convert the force-moment tensor above to the stress tensor in fourier space}
The force-moment tensor in Fourier space is given by
\begin{equation}\label{FMtens}
    \hat{\sigma}(\vec{q}) = \sum_{i=1}^{N} \hat{\sigma}_i \exp(i\vec{q} \cdot \vec{r}_i)
\end{equation}
 To obtain the stress tensor in Fourier space we need to divide Eq.~\ref{FMtens} by the volume (area) $V$ of the simulation box, however, for simplicity, we ignore the constant factor of $1/V$ in our calculations. 
The stress correlations in Fourier space is computed using:
\begin{eqnarray}\label{formulacorel}
C_{klmn} (\vec{q}) = \langle \Delta \sigma_{kl}(\vec{q}) \Delta \sigma_{mn}(-\vec{q})\rangle \\
\Delta \sigma_{kl}(\vec{q}) = \sum_{i=1}^{N} (\sigma_{kl,i} - \langle \sigma_{kl} \rangle) \exp(i\vec{q} \cdot\vec{r}_i)
\end{eqnarray}
where $\langle \sigma_{kl} \rangle$ is the average value of the stress component. The angular brackets $\langle ...\rangle$ denotes an ensemble average performed over all the particles and the average over all statistically independent samples prepared in the same conditions. For 3$D$ gels, the computation of $\hat{\sigma}_i$ also includes the contribution from the three body or angular term of the interaction potential \cite{thompson2009general}.

From the stress correlation calculations, we construct maps of the correlation intensity in Fourier space in polar $(q,\theta)$ representation for 2$D$ gels, for a range of $q$ from $2\pi/L$ to $56\pi/L$. For the angular variation of the correlations, along with the radial averaging, the data are also averaged over a small angular bin width of $\approx 6^{\circ}$ to reduce noise. For comparison, we also compute the correlations in real space along both $x$ and $y$ direction with the coarse-graining length $d$.    
%{\color{red} BC comment Again, averaged over some radial $q$, correct ?We should mention this range of $q$.},
In 3D, the range of $q$ used is from $2\pi/L$ to $42\pi/L$, the stress correlations are averaged over $|q|$ and projected in 2D using the Hammer projection coordinate system $(H_x,H_y)$\cite{nampoothiri2020emergent,nampoothiri2022tensor}. $H_x$ and $H_y$ are computed using
\[
H_x = \frac{2\sqrt{2} \cos{\alpha} \sin{\beta/2}}{\sqrt{1+\cos{\alpha}\cos{\beta/2}}}
\]
\[
H_y = \frac{\sqrt{2} \sin{\alpha}}{\sqrt{1+\cos{\alpha}\cos{\beta/2}}}
\]
where $\alpha$ and $\beta$ are the latitude and longitude, respectively. $\alpha = \theta - \pi/2$ and $\beta = \varphi$, where $(\theta,\varphi)$ are the spherical polar coordinate angles.

All stress correlations are scaled by the maximum value of the $C_{xxxx}$ correlation function.
\begin{figure}[ht!]
%\centering
\includegraphics[scale=0.9]{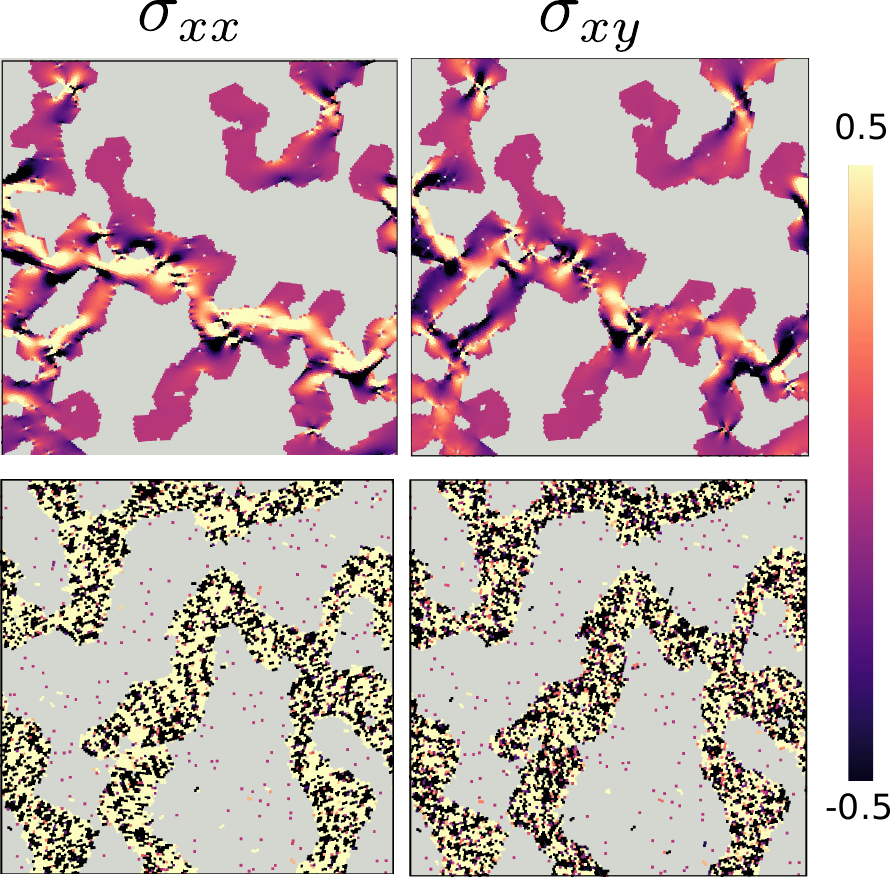}
\caption{\label{stress2D} Snapshots of 2D gel showing $xx$ and $xy$ components of the particle stresses for a rigid (top row) and floppy gel (bottom row).}
\end{figure}
\begin{figure}[ht!]
%\centering
\includegraphics[scale=0.82]{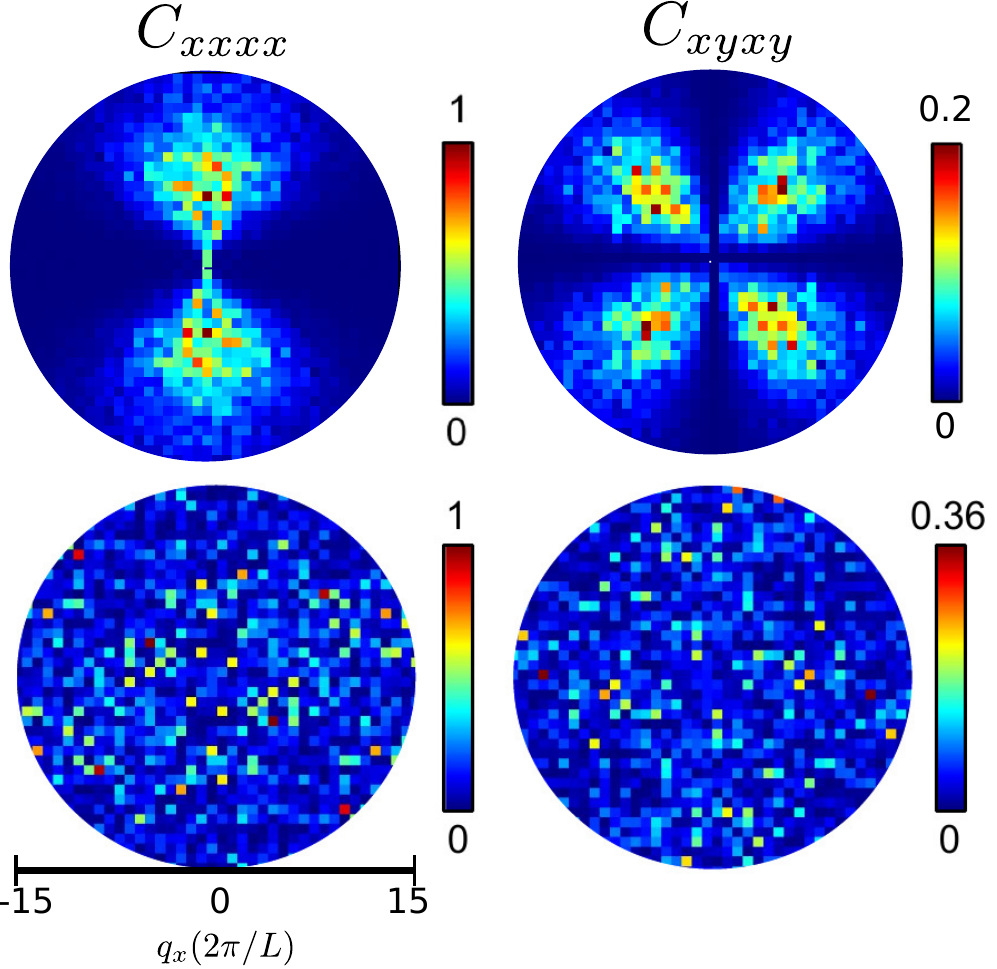}
\caption{\label{corel2D} Stress correlations $C_{xxxx}$ and $C_{xyxy}$ in Fourier space for the rigid and floppy gel in the top and bottom row, respectively.}
\end{figure}
\begin{figure*}[ht!]
%\centering
\includegraphics[scale=0.9]{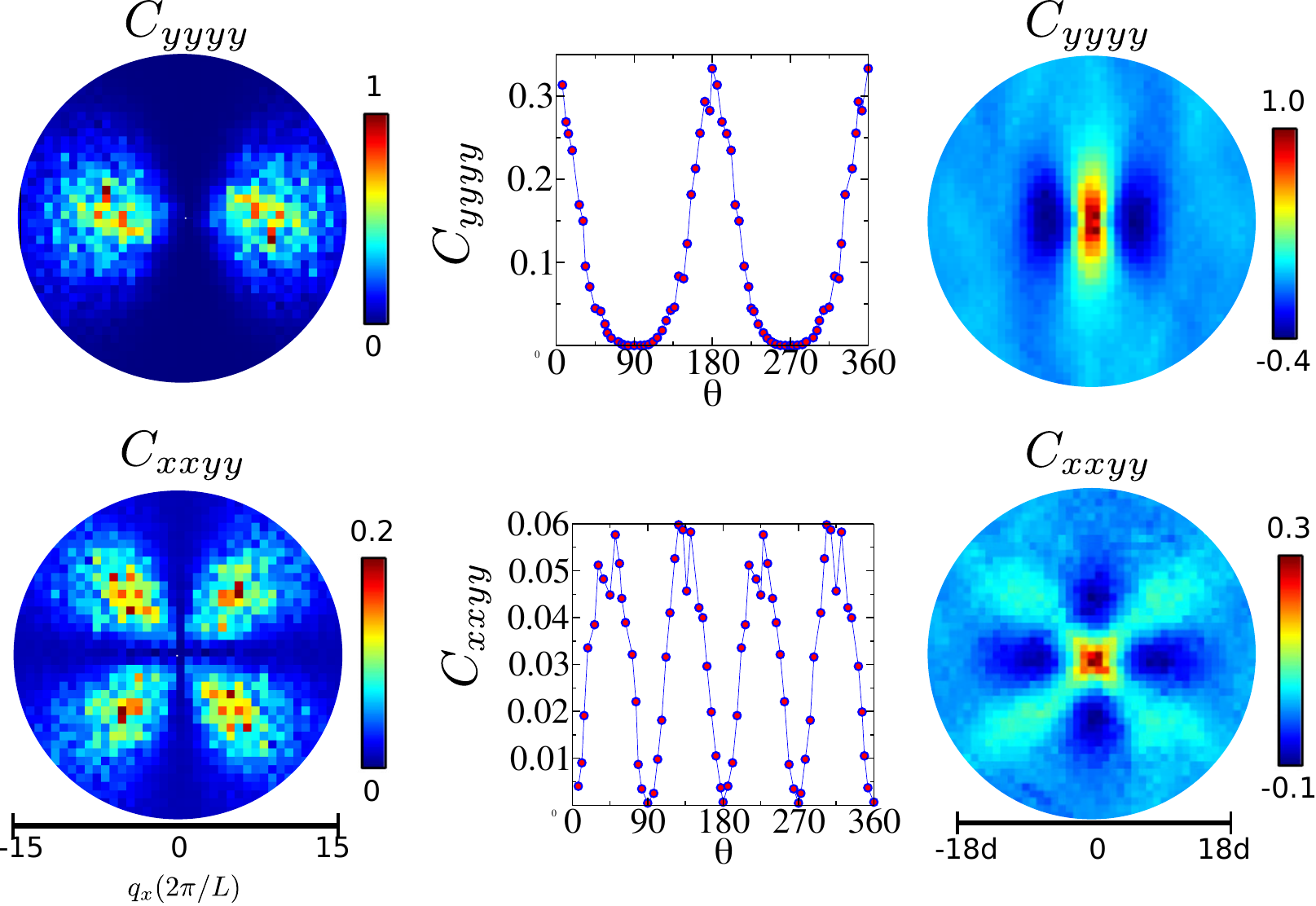}
\caption{\label{2Dgelcorels} Stress correlations $C_{yyyy}$ (top row) and $C_{xxyy}$ (bottom row) in $q$- space (left panel) and real space (right panel). In the middle panel the angular dependence of the $q$- space correlations is shown. }
\end{figure*}
\begin{figure}[ht!]
%\centering
\includegraphics[scale=1.0]{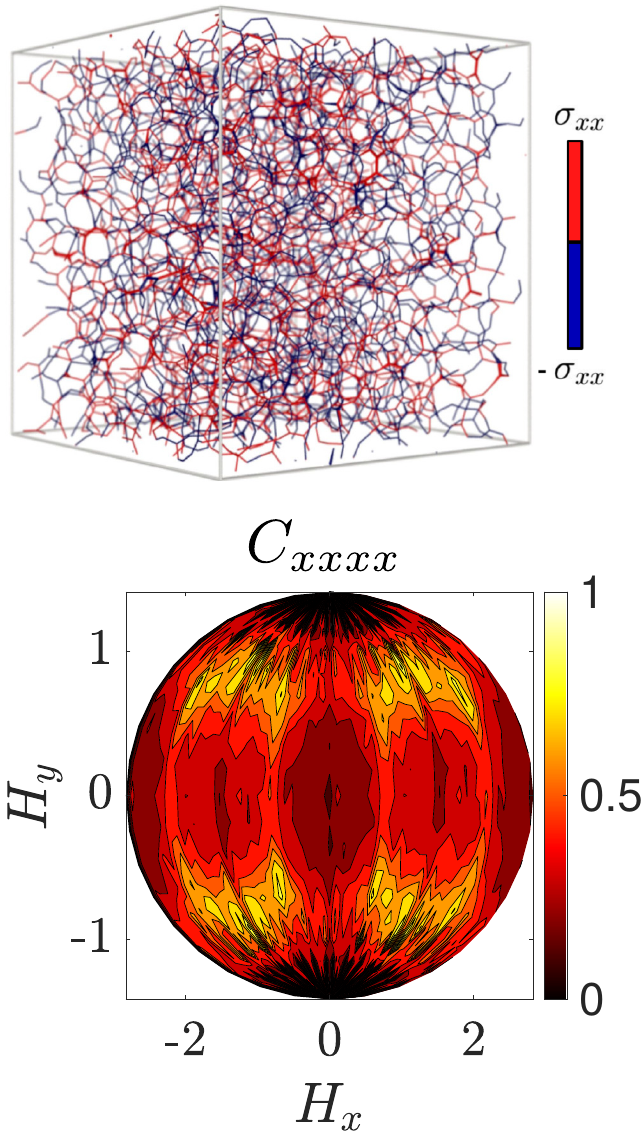}
\caption{\label{3dcorelxx} (Top panel) Snapshot of the gel network at $\phi=0.125$, only the bonds are shown. The particle bonds are red if the particle stress is positive and blue if the particle stress is negative. The bond width is proportional to the magnitude of the stress. (Bottom panel) Hammer projection of $C_{xxxx}$ in Fourier space of the 3D gels at $\phi=0.125$.}
\end{figure}
\begin{figure}[ht!]
%\centering
\includegraphics[scale=1.2]{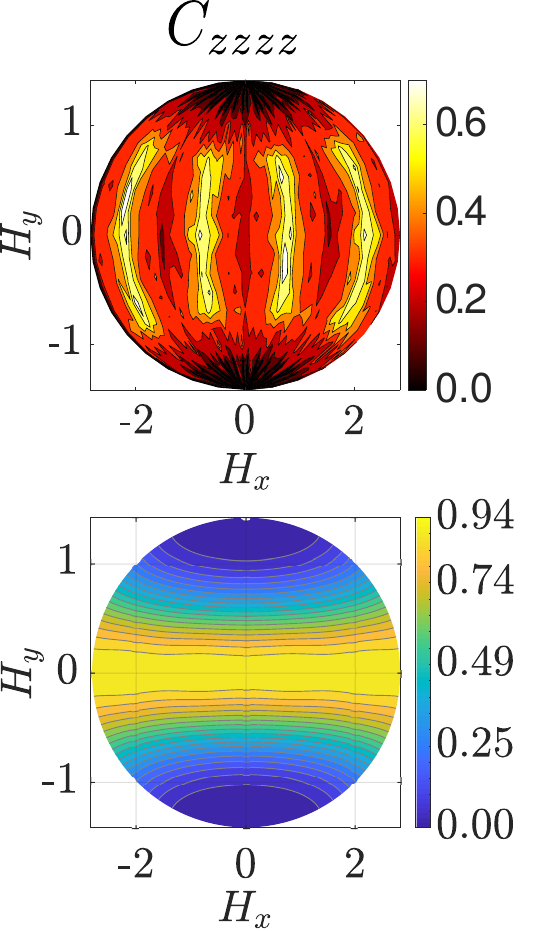}
\caption{\label{3dcorelzz} Comparison of $C_{zzzz}$ of gels at $\phi=0.125$ (top figure) with the isotropic jammed solids (bottom figure). The jammed solids data is from Ref. \cite{nampoothiri2022tensor}}
\end{figure}
\begin{figure*}[ht!]
%\centering
\includegraphics[scale=0.8]{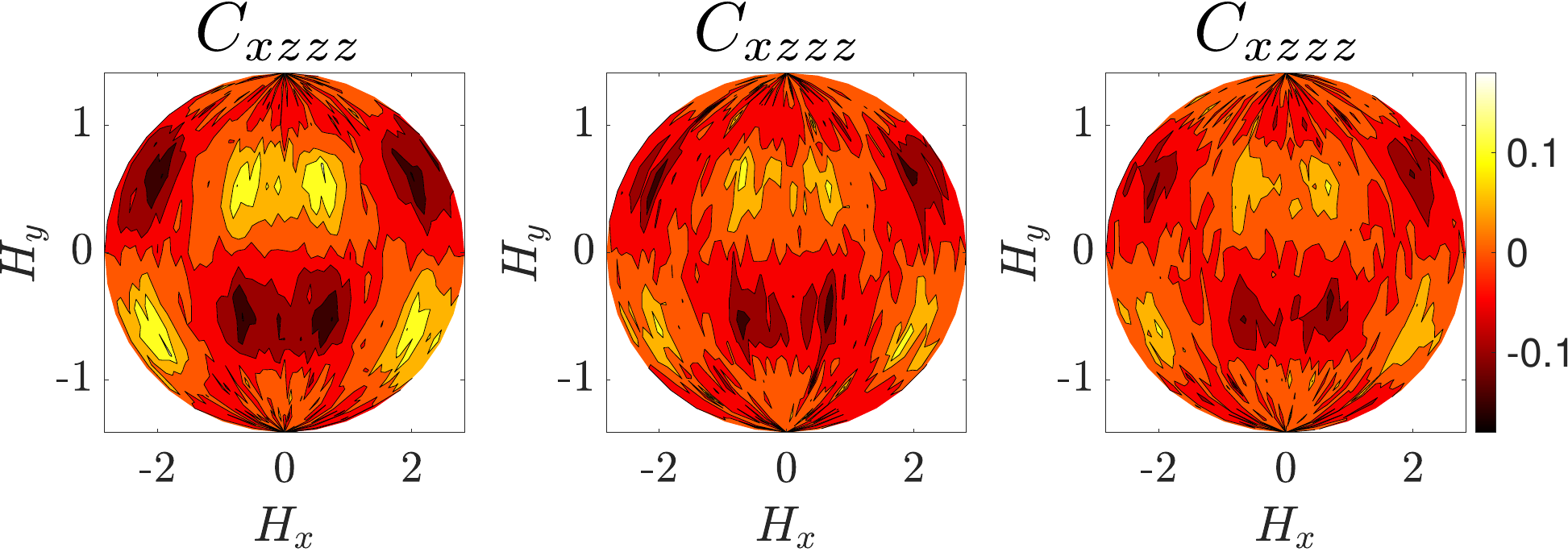}
\caption{\label{3dcorelphis} Hammer projection of $C_{xzzz}$ for the gel configurations at packing fraction $0.125$ (left), $0.075$ (middle) and $0.05$ (right). The gel configurations are prepared with cooling rate $C_r = 9 \times 10^{-5}\epsilon/(\tau_{0} k_{B})$.}
\end{figure*}

\section{Results}
\label{results}
We now discuss the result obtained for 2$D$ and 3$D$ model particulate gels in mechanical equilibrium.
%In this section, we present the stress correlation data for 2D and 3D gels in mechanical equilibrium.
As discussed in the Introduction, soft particulate gels have strong structural heterogeneities which depend on the aggregation kinetics and the path to gelation. These structural heterogeneities are associated to stress heterogeneities, however the microstructural origin of the stress heterogeneities is not easily identified. Fig. \ref{stress2D} shows snapshots of a rigid gel close to the rigidity threshold \cite{zhang2019correlated} (top panels) and a floppy gel (bottom panel) from our $2D$ simulations. The particles are colored according to their magnitude of $xx$ and $xy$ component of $\hat{\sigma}_{i}$ computed using Eq. \ref{FMpar}. For these numerical samples, these components of the stress tensor have a magnitude varying between $-10 \epsilon/d^{3}$ and $8 \epsilon/d^{3}$. If we use the whole range for $\sigma_{xx}$ and $\sigma_{xy}$ in the color maps the gels appear mechanically homogeneous, as they are overall soft. By zooming in the stress variation, down to $\Delta \sigma_{ij} \simeq 1 \epsilon/d^{3}$, chain-like patterns emerge in the stress maps of rigid gels. Floppy gels feature relatively large stress fluctuations, even at this large stress magnification, which however appear randomly distributed and uncorrelated. For rigid gels, instead, the stress fluctuations display an anisotropic pattern, which strongly suggests anisotropic and localized stress correlation across the samples and is reminiscent of force chains in granular packings \cite{cates1998jamming,radjai1998bimodal,gendelman2016determines,behringer2018physics,vinutha2019force,wang2020connecting}. 
%the high and low-stress regions form a dipole and are localized at joints where the small clusters attach to the spanning network during gel formation. 
%However with a lower value of the stress threshold we observe stress chains for $\sigma_{xx}$ and $\sigma_{xy}$ maps in the rigid case. In the floppy case, we see a lot of rattler particles that are not a part of the spanning network and there is no stress localization or pattern.

These qualitative observations can be quantitatively confirmed by analyzing the intensity maps of the stress correlations, for the corresponding component of the stress tensor, as shown in Figure \ref{corel2D} (left). While for floppy gels the stress-stress correlation maps do not feature any distinctive pattern, for rigid gels they do. The patterns indicate both the presence of correlations that span the whole system, and the fact that they do so in a very directional and anisotropic fashion, in spite of the particle interactions and the overall gels being isotropic. 
The $C_{xxxx}$ map obtained from the rigid gels shows
%{\color{red} comment about what the shape of the pattern means}. 
a large $q$ cutoff in the correlations, which simply reflects the fact that well inside the gel branches the material is spatially and mechanically homogeneous. However the correlation intensity is significant (along the $q_x$ axis) even for the smallest $q$, i.e., for distances up to the whole system size. The value of  $C_{xxxx}$ being instead zero along the line $q_y = 0$ indicates that the forces can propagate approximately only along their own direction and very unlikely perpendicular to it. This is a direct consequence of force balance, as discussed in Section \ref{theorypred} and provides a quantitative evidence to the presence of force chains. The same characteristics, i.e., the long range correlations and the strong anisotropy indicated by the angular dependence, are also present in the four-lobe pattern of the $C_{xyxy}$ map for the rigid gels (Fig. \ref{corel2D} right). We note that for $C_{xyxy}$ the signal at low $q$ is much weaker, probably due to system size effects and the proximity to the rigidity threshold of the rigid gels considered here. %seem to have a small $q$-cutoff because the system is close to the boundary of rigidity threshold and most of the configurations have spanning rigid cluster only along one direction. The small $q$-cutoff is due finite size effects. 
%The angular dependence of the correlations contains the information of the gel modulus ($\Lambda^{-1}$ tensor). The $C_{xyxy}$ and $C_{yyyy}$ correlations seem to have a small $q$-cutoff because the system is close to the boundary of rigidity threshold and most of the configurations have spanning rigid cluster only along one direction. The small $q$-cutoff is due finite size effects.  

The two-lobe pattern in $C_{xxxx}$ and four-lobe pattern in $C_{xyxy}$ for the rigid gels are identical to the correlation patterns already obtained in experiments and simulations of granular solids, confirming the pinch-point structure of stress correlations in Fourier space predicted by the elasticity theory of the VCT framework \cite{nampoothiri2020emergent}. In Fig. \ref{2Dgelcorels}, we also show the $C_{yyyy}$ (top) and $C_{xxyy}$ (bottom) maps for the 2D rigid gels in Fourier space (left) and the corresponding radially averaged angular plots (center). The stress correlations maps in the real space (right) complement, and confirm, the insight gained with the maps of the correlations in Fourier space. The remaining stress correlations for 2D gels, considering all stress tensor components, are shown in Appendix A. These maps quantitatively establish the similarity of stress transmission in soft gels and in granular solids. They support the idea that the rigidity and elasticity of soft particulate gels, when particle interactions are sufficiently large with respect to $k_{B}T$, can be fundamentally understood as an emerging property which is the result of the local and global constraints imposed by mechanical equilibrium, rather than just a consequence of their microscopic interactions. We note that, as briefly discussed in Section \ref{theory-simulations} A (see eq.\ref{eq:VCTG_field_eqns}), in the VCT framework the angular dependence of the correlation intensity (Fig. \ref{2Dgelcorels} center) provides direct information on the properties of the emerging elastic tensor $\Lambda$. When comparing our data to the results in Ref.\cite{nampoothiri2020emergent}, we notice interesting quantitative differences in the corresponding angular plots, therefore suggesting that the obvious significant differences between the overall elastic behavior of gels and granular solids, can also be further investigated with this approach. 

%Furthermore, a more detailed comparison across the correlation patterns for all stress components between gels and granular solids can provide novel insight into    

%Gels differ from the granular solids in two aspects: first, the density heterogeneity, and second, the gel networks can sustain both compression and tension. The long-ranged anisotropic nature of the stress correlations and the pinch point singularities in rigid gels are similar to the granular solids. It's interesting that the correlation patterns appear to be universal to amorphous solids. The specific pattern of the stress correlations is a consequence of the self-organization of particles to form a rigid structure. \\
\begin{figure}[ht!]
%\centering
\includegraphics[scale=0.8]{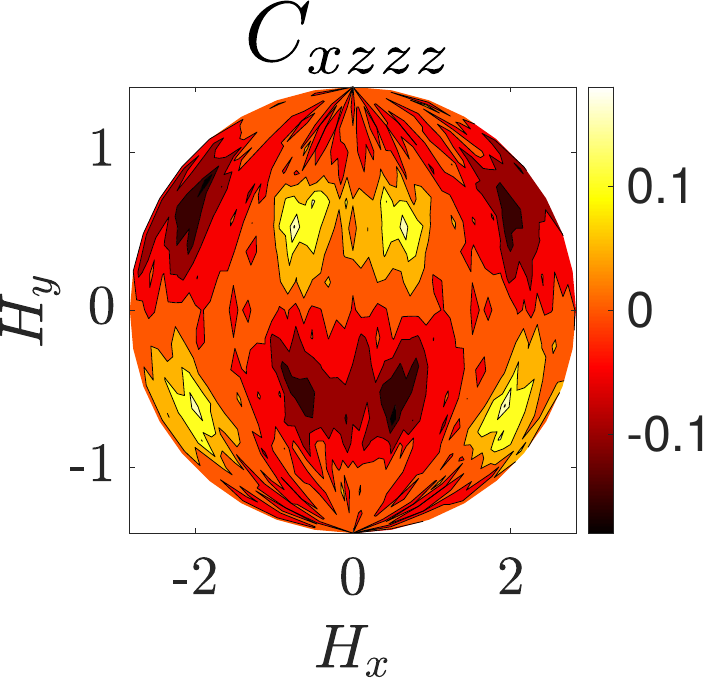}
\caption{\label{3dcrs} Hammer projection of $C_{xzzz}$ for the gel configurations at $\phi=0.125$ and $C_r = 9 \times 10^{-2} \epsilon/(\tau_{0} k_{B})$.}
\end{figure}
In the case of 3$D$ gels, we have computed all $21$ stress-stress correlations. We discuss some of them here in detail, while the remaining ones are shown in Appendix B. A snapshot of one of the gel networks (at volume fraction $\phi=0.125$ and obtained with cooling rate $C_{r}=  9 \times 10^{-5}\epsilon/(\tau_{0} k_{B})$ is shown in Fig. \ref{3dcorelxx} (top panel). We have colored the particle bonds in red if the $xx$ component of the particle stress is positive (tension) and blue if negative (compression), to highlight how, even with this information, the presence of force chains or stress localization can not be easily detected. However, the maps of the corresponding stress correlation function clearly reveals their presence: also in this case all stress-stress correlations are long-ranged, as determined by the mechanical equilibrium constraints \cite{henkes2009statistical,lemaitre2018stress}, and the $2D$ Hammer projection of the correlation intensity for the fluctuations of the $xx$ component of the stress tensor (Fig.~\ref{3dcorelxx}, bottom panel) are strongly anisotropic, with the angular pattern indicating that stress transmit across the sample along specific and localized directions. 

In Figure \ref{3dcorelzz}, we compare the intensity pattern of $C_{zzzz}$ in rigid gels (top) and granular solids (bottom, from Ref. \cite{nampoothiri2022tensor}) to point out the striking difference in the angular dependence. The map obtained for a model granular solid shows that the stress-stress correlations are basically independent on the azimuthal angle $\phi$ and only depend on $\theta$, whereas the pattern for the gels indicates a strong dependence on both $\theta$ and $\phi$, confirming that this approach is able not just to reveal similarities and common traits, but also to identify differences. With this respect, we think that this striking difference in the angular dependence may be related to the fact that all structures in a granular solid can only transmit compression or shear stresses, whereas in gels also tension can be transmitted. %   The difference may be due to the bending rigidity term in the interaction potential that favors particular bond angles or that the gel networks can sustain both the compressive and tensile stresses.  
All the remaining stress correlations from $3D$ gels at $\phi=0.125$ are shown in Appendix B.

The pattern of the stress correlation in gels are sensitive to the distance from the rigidity threshold, as shown in Fig. \ref{3dcorelphis}, where the $C_{xzzz}$ maps are displayed for gels at different $\phi = 0.125,0.075$ and $0.05$ obtained for cooling rate $C_r = 9 \times 10^{-5}\epsilon/(\tau_{0} k_{B})$. As the volume fraction increases the local connectivity and gel moduli increase. At $\phi=0.05$, the networks are sparse, spatially heterogeneous and barely rigid, as also discussed in Refs. \cite{bantawa2022, bouzid2018network}, and their elasticity is extremely week. With increasing the particle volume fraction, the gels obtained with the same cooling rate become stiffer, more locally connected and move away from the rigidity threshold. The stress correlations at different densities, interestingly, have the same general pattern, which however becomes increasingly blurred as the gels approach the rigidity threshold (Fig. \ref{3dcorelphis} from left to right). A similar trend is observed for the correlations of all stress components. These findings support the idea that the stress correlation intensity variation is directly related to the elastic moduli, suggesting that they may be used to detect the distance from the rigidity threshold and provide information on the marginal stability of the gels. We confirm these findings also by varying, for the same particle volume fraction, the cooling rate at which the gel is formed, since increaseing this rate leads to gels that are weaker and closer to the rigidity threshold. By comparing, for example, the stress correlation map in Fig. \ref{3dcrs}, which shows $C_{xzzz}$ for $\phi=0.125$ at $C_r = 9 \times 10^{-2}\epsilon/(\tau_{0} k_{B})$, with the first map on the left of Figure \ref{3dcorelphis} (corresponding to the same volume fraction and $C_r = 9 \times 10^{-5}\epsilon/(\tau_{0} k_{B})$) we can notice how again the correlation pattern becomes more blurred for weaker and more marginal gels. Clearly, therefore, the changes seen in the stress correlations are due to the different elasticity of the network and not simply the gel density.
\section{Summary}
\label{conclu}
In summary, we have demonstrated the presence of force chains in soft particulate gels by computing the stress-stress correlation functions in 2D and 3D model gels. The stress correlation patterns distinguish the rigid gels from the floppy gels and display the same general characteristics as granular solids, i.e., the strong angular dependence, and the pinch point singularity predicted by a stress-only theoretical framework (the VCT framework in Ref. \cite{nampoothiri2020emergent}), where the elastic response is a property emerging only from the constraints of mechanical equilibrium, without resorting to the constitutive relation of linear elasticity. We also demonstrate that the stress-stress correlation patterns nevertheless highlight distinctive differences between gels and granular solids, and that they are sensitive to variation of the gel moduli, due to changes in the gel topology and marginal stability, not simply to the gel density. These results are consistent with the idea that the stress-stress correlation patterns can provide further information on the tensorial properties of the emerging elasticity in these materials. Stress-stress correlations patterns have been measured in experiments on granular solids using photoelastic disks (see Ref.~\cite{nampoothiri2020emergent}), however this is clearly much more challenging for experiments on particulate gels, where measurement techniques to extract local stresses and their spatial distributions are still being developed \cite{arevalo2015stress,Lin2016relating,gauthier2021new,dong2022direct}. Using the stress-stress correlation functions obtained in simulations to extract, from their angular dependence, the properties of the elastic tensor, instead, would provide predictions for the overall mechanical response to different deformation modes (e.g. shear, compression, extension, etc...) which are indeed accessible to mechanical and rheological experiments. This information, testable on a broad range of gels and experimental conditions, could then be fed back to the theory to obtain the stress-stress correlation patterns from the experiments. Hence developing further the analysis proposed here seems a promising route to gain novel insight into the emerging elasticity of soft particulate gels and its connection to stress localization.\\

%%%%%%%
\begin{acknowledgments}
We thank Shang Zhang, Jishnu N Nampoothiri and Minaspi Bantawa for valuable discussions and configurations. This work was supported by the National Science Foundation (NSF-DMR-2026842; NSF-DMR-2026825 and NSF-DMR-2026834) and the Clare Boothe Luce Program. Acknowledgement is made to the donors of the American Chemical Society Petroleum Research Fund for partial support of this research.
\end{acknowledgments}
%\section*{Data availability}
%The data that support the findings of this study are openly available in the University repository at
\nocite{*}
\bibliography{forcechain_gels_v1.bib}
\clearpage
\onecolumngrid
\appendix
\section{2D correlation functions}
 In Fig. \ref{2Dgelcorels}, we show data for $C_{yyyy}$ and $C_{xxyy}$. Here, we show all the remaining stress correlations for the rigid gels. The correlation functions are computed using Eq. \ref{formulacorel}. The angular dependence of these correlation functions matches with the granular case \cite{nampoothiri2020emergent}.
\begin{figure*}[h!]
%\centering
\includegraphics[scale=0.9]{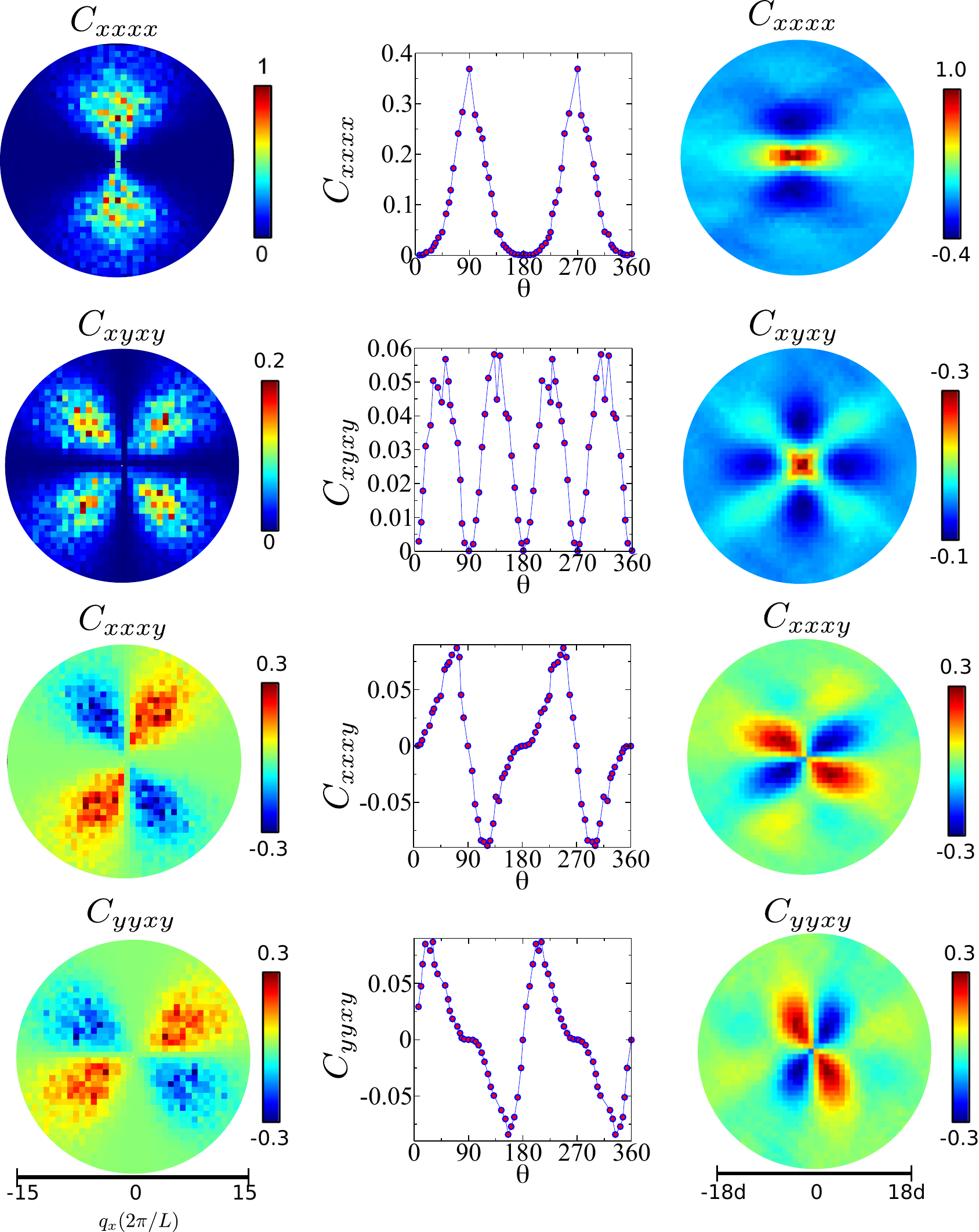}
\caption{\label{2Dcorelsall} The stress correlation functions for the 2D rigid gels. The three columns show respectively: the stress correlation in $q-$ space, and the corresponding radially averaged angular plots and the stress correlation in real space.}
\end{figure*}
\section{Stress correlations in Fourier space for 3D gels}
\begin{figure*}[h!]
\includegraphics[scale=1.1]{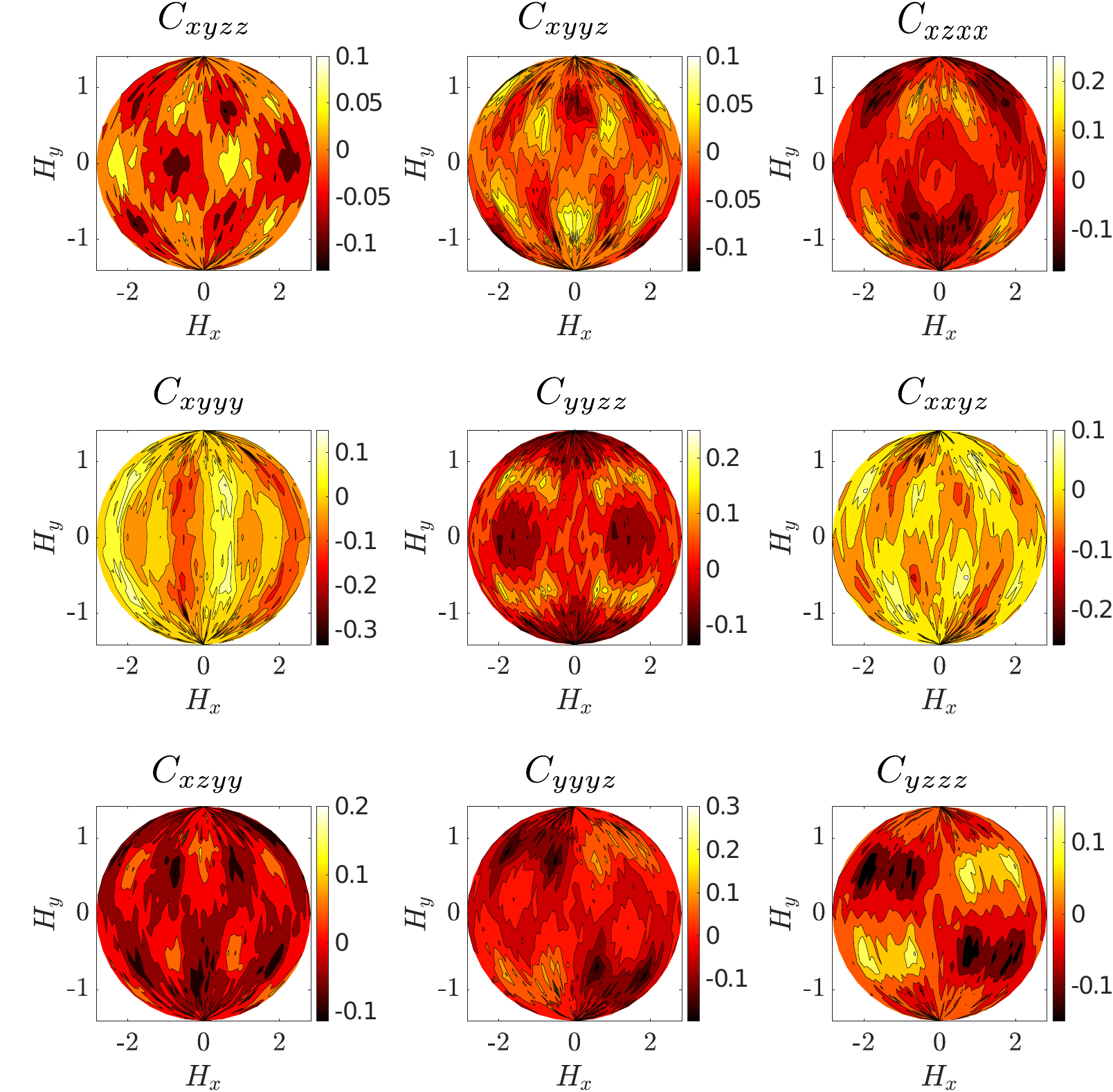}
\caption{\label{3Dcorels} The $q$-space stress correlations for 3D gels at $\phi=0.125$.}
\end{figure*}

In the manuscript, we show $C_{xxxx}, C_{zzzz}$ and $C_{xzzz}$ correlation functions of gels at $\phi=0.125$. In Fig. \ref{3Dcorels}, the remaining $18$ correlation functions are shown.
\begin{figure*}[h!]
\includegraphics[scale=1.1]{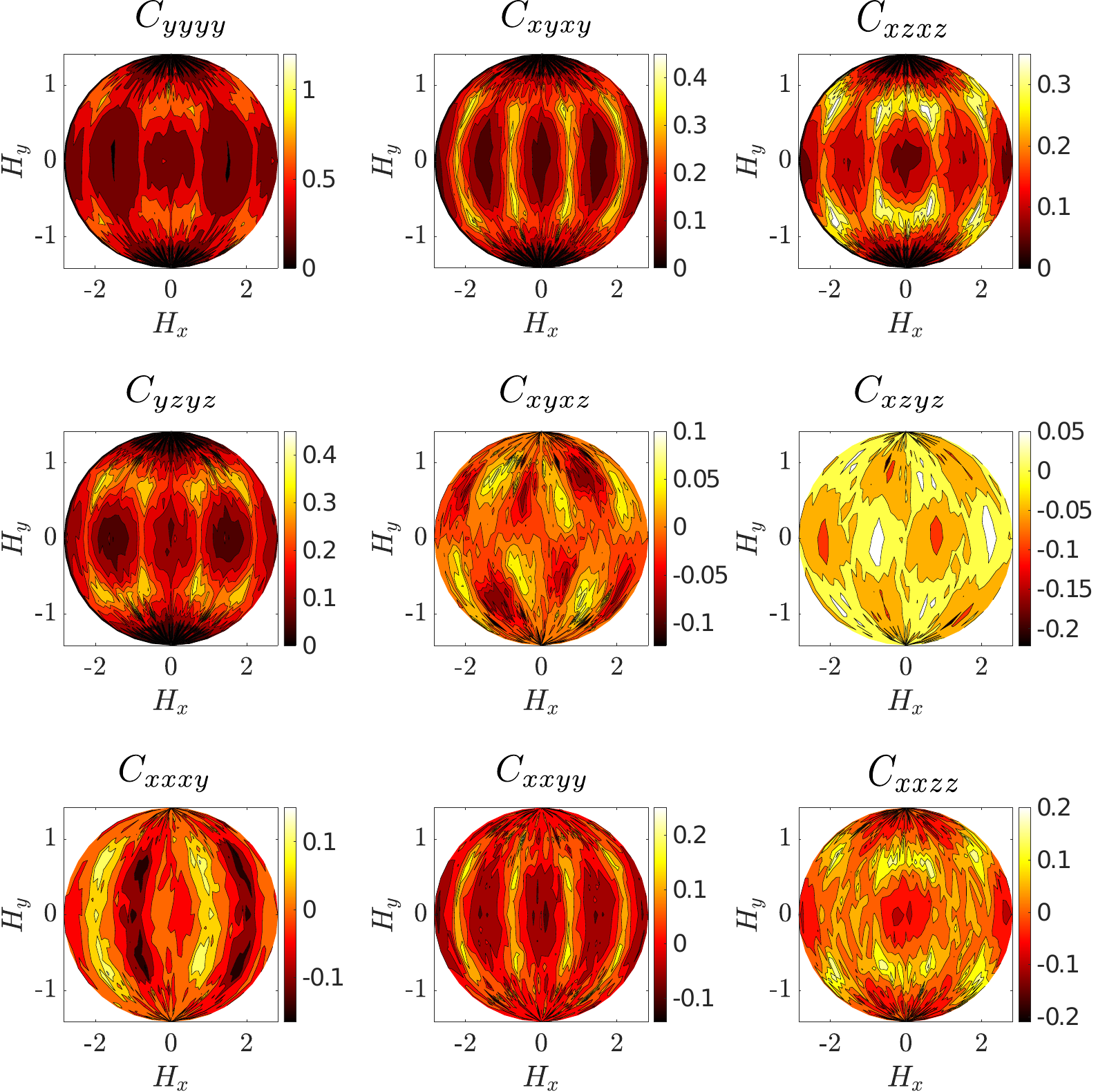}
\end{figure*}

\end{document}